\documentstyle[aps]{revtex}
\preprint{HEP/123-qed}
\draft
\def\bea{\begin{eqnarray}}
\def\eea{\end{eqnarray}}
\def\bfg{\begin{figure}}
\def\efg{\end{figure}}
\def\be{\begin{equation}}
\def\ee{\end{equation}}
\def\bn{\begin{enumerate}}
\def\en{\end{enumerate}}
\def\bi{\begin{itemize}}
\def\ei{\end{itemize}}
\def\ba{\begin{array}}
\def\ea{\end{array}}
\def\bm{\begin{mathletters}}
\def\em{\end{mathletters}}

\def\d{{\rm d}}
\def\dt{{\rm d}t}
\def\wide{\widetext\noindent}
\def\narrow{\narrowtext\noindent}
\def\nonum{\nonumber\\}
\date{\today}
\begin{document}
\title{Decoherence-induced wave packet splitting} 
\author{Young-Tak Chough, Hyunchul Nha, Sang Wook Kim, Sun-Hyun Youn$^\dag$ and Kyungwon An}
\address{Center for Macroscopic Quantum-field Lasers, Department of Physics\\ 
Korea Advanced Institute of Science and Technology, Taejeon 305-701, Korea\\
$^\dag$Department of Physics, Chonnam National University, Kwangju, Korea}
\maketitle

\begin{abstract}
We provide an intuitive interpretation of the optical Stern-Gerlach effect (OSGE) in the dressed-state 
point of view. We also analyze the effect of atomic damping in an experiment on the OSGE. We show that 
the atomic damping also causes the wave packet splitting, in a {\it non-mechanical} fashion, as opposed 
to the coherent process that is {\it mechanical}.
\end{abstract}
\narrowtext

\section{Introduction}

The mechanical interaction of light and matter has been comprehensively studied since the pioneering 
work of Kapitza and Dirac in 1933\cite{KDE}. Much work has been done on the atomic beam deflection, 
diffraction, refraction or interference\cite{meystre,gould,rempe,shore,atom interference} by a 
standing-wave field, classical or quantized, and recent achievement of optical cooling and trapping 
of neutral atoms\cite{cooling} using the light pressure force may be said to have a root in these 
themes. One of the interesting issues that came up in the theme of atom-field mechanical interaction 
is the so-called optical Stern-Gerlach effect (OSGE)\cite{cook,tanguy,kazantsev,sleator}. It has been 
explored at various levels of sophistication since mid seventies when it was suggested that the 
trajectory of a two-state atom interacting with an optical field gradient can be split into two paths, 
each path containing atoms in one of the two orthogonal dressed states, under certain 
circumstances\cite{kazantsev}. It is thus an optical analogy of the well-known magnetic Stern-Gerlach 
effect (MSGE)\cite{MSGE} in which the trajectory of a spin-$1\over2$ particle is split into two paths 
in a magnetic field gradient. %Hence the name OSGE.

The OSGE was experimentally demonstrated in 1992 by Sleator and others in the near-infrared $(\lambda
\sim1\,\mu$m) with metastable helium atoms (He$^\ast$)\cite{sleator}. The experimental result seems to 
be in agreement with the semiclassical theories\cite{cook,tanguy,kazantsev}, but later it was also 
pointed out that the OSGE may show some additional peculiarities, sensitively depending on the quantum 
nature of the field\cite{mss}. Not only incorporating the quantum nature of light and the atomic 
center-of-mass motion in the picture, recent works also take into consideration the finite spatial 
extent of the atomic wave packet\cite{vaglica,ishkhanyan,tak}, rather than treating an atom as a point 
mass\cite{gould} or a plane wave\cite{cook,tanguy} in space. Particularly, the first reference 
in\cite{vaglica} presents quite an elegant operator method in such a full quantum-physical description 
of the OSGE.

Noting, however, that the theories originally envisaged the OSGE as the coherent processes dealing 
with ideal lossless systems, mostly in the linear regime, we explored the role of the decoherent 
processes in the OSGE, particularly in the nonlinear regime, which may be significant in the optical 
frequencies. For the systems free of damping, decent analytic treatments may be available, but when the 
system is open to its environment so that the coherent dynamical evolution of the system is frequently 
interrupted by the discontinuous processes such as atomic spontaneous decays or cavity emissions, etc., 
analytic approaches may not be always possible. There are a number of theoretical methods in dealing 
with such open quantum systems, but we will resort to the formalism of the quantum trajectory theory 
(QTT)\cite{qtt} particularly because the theory provides most direct and intuitive interpretations on 
the behavior of the system we look into. In particular, we will show in the viewpoint of QTT that the 
atomic damping can also cause the wave packet splitting, and that in a {\it non-mechanical} fashion, 
as opposed to the coherent process that is {\it mechanical}.

In Sect.\,\ref{sect2}, we first provide our simple and intuitive illustration on the OSGE, based on 
the dressed-state picture\cite{dressed state}, and in Sect.\,\ref{sect3}, we discuss the effect of 
damping in the language of QTT. In Sect.\,\ref{sect4}, we perform a numerical simulation of a typical 
experiment and compare the outcomes of such a realistic experiment with the predictions made for the
ideal systems free of damping. Sect.\,\ref{sect5} summarizes this work.
 
\section{The optical Stern-Gerlach effect}
\label{sect2}

\subsection{The wave packet ``pulsation''}

Consider a simple model in which a Gaussian wave packet of a two-state atom initially in its ground 
state is placed on a node of a standing wave cavity, on the cavity axis, as illustrated in 
FIG.\ref{model}. In order to quickly appreciate the essential physics, let us assume that the cavity 
field is initially in a single-quantum Fock state, and the atom and the cavity are on resonance. We 
will neglect damping for the time being. The interaction Hamiltonian is then simply written as 
\be
\label{hamiltonian}
H={p^2\over2M}+i\hbar g(x)\left(\sigma_-a^\dag-\sigma_+a\right)
\ee
where $p$ is the atomic momentum operator conjugate to the position $x$ on the cavity axis; $M$, the 
atomic mass; and $g(x)$, the atom-field coupling strength at position $x$ such that 
\be
 g(x)=g_0\sin kx\,,
\ee
$g_0$ being a constant and $k$ the field wave number. $a^\dag(a)$ is the creation(annihilation) 
operator for the field, and $\sigma_+=|e\rangle\langle g|=(\sigma_-)^\dag$ is the atomic excitation 
operator, with $e(g)$ labeling the atomic excited(ground) state. 

We will focus only on a spatially well-localized---i.e., a particle-like---wave packet such that its 
initial width $\Delta x$ is much smaller than the wavelength $\lambda$ of the cavity field. We choose 
$\Delta x=\lambda/40\pi$ which corresponds to the width in the momentum space $\Delta p=10\hbar k$. 
Due to the finite spatial extent of the wave packet, the atom-field coupling strength at the node is 
nonzero, although very small compared to its peak value $g_0$, and the interaction slowly commences. 
Note that the initial effective coupling strength is given by an overlap integral of the mode function 
and the atomic spatial probability distribution, i.e.,
\bea
\label{ue}
 g_{\rm eff}&=&\int\!\d x\,\psi_0^2(x)\left|g(x)\right|\nonum
                   &=&2\int_0^\infty\!\!{1\over\sqrt{2\pi(\Delta x)^2}}
\exp\left[-{x^2\over2(\Delta x)^2}\right]g_0\sin kx\,\d x\nonum
            &\sim&2\sqrt{2\pi}\left({\Delta x\over\lambda}\right)g_0\sim0.04g_0
\eea
where $\psi_0(x)$ is the initial atomic wave packet and we put $\sin kx\sim kx$ as $\Delta x\ll\lambda$. 
Thus the effective coupling strength would be only about 4\,\% of $g_0$ initially. 

Let us first present our numerical results on the evolution of the atomic wave packet in 
FIGs.\,\ref{Px-Fock}(a) and (b). Note that since the atomic mass $M$, field wave number $k$ and 
atom-field coupling constant $g_0$ appear in the Hamiltonian at the same time, one needs to specify 
these quantities somehow, but the only system-dependent parameter is the dimensionless factor 
$\mu=\hbar k^2/2Mg_0$, which appears from the Schr\"odinger equation when we rescale time in the unit 
of $g_0^{-1}$ and the momentum in $\hbar k$. [Note that $\mu=k(\hbar k/M)/2g_0=kv_1/2g_0$ is the ratio 
of the Doppler shift by the single photon recoil to the single photon Rabi frequency.] Let us take 
$\mu=1.7\times10^{-4}$ (with $M=100$\,a.u., $\lambda=600$\,nm, and $g_0=200$\,MHz) for a typical atom 
interacting with an optical field in the strong coupling regime. 

The figures show that the initial Gaussian packet evolves into a double-peaked distribution and then 
back to its original shape, as time proceeds. The inset is a ground-level view of the packet in the 
coordinate space along with the $\cos(kx)$ curve, which clearly shows that the two split bumps make 
turns exactly at the neighboring nodes. Let us first note that the two figures are {\it mechanically} 
consistent in the following sense. Let $\tau=g_0t$. In the momentum space, the position $\bar p$ of 
one of the split peaks grows roughly linearly in time in the fashion, $\bar p(\tau)\sim0.9\hbar k\tau$ 
until $\tau\sim100$, and then bounces back to the original position. So, the position $\bar x$ of a 
peak after one cycle of motion in the momentum space is roughly
\be
 \label{x-estimation}
 \bar x\sim2\int_0^{100/g_0}{\bar p\over M}\dt\sim(2.9\times10^3\mu)\lambda=0.49\lambda
 \sim0.5\lambda
\ee
which is in a good agreement with FIG.\,\ref{Px-Fock}(b). Although this wave packet motion can be said 
to have essentially the quantum-mechanical origin in the sense that $\bar x$ vanishes if $\hbar$ does, 
the motion conforms at the same time to the classical-mechanics. We will get back to this 
apparently trivial observation when we include damping in the system in Sect.\,\ref{sect2}, and will 
point out some interesting features in connection to this issue.    

Thus we have seen an exotic phenomenon of {\it wave packet pulsation on a node}. Note that the analytic 
approaches such as\cite{vaglica} based upon the linear approximation did not predict this 
{\it pulsation}. The descriptions are mostly limited to the wave packet splitting in an early stage of 
the atom-field interaction. If the atoms are let to leave the interaction region at some point of time, 
the emerging wave packet will have a pair of distinguishably separated probability density peaks. This 
phenomenon of wave packet splitting is the very optical Stern-Gerlach effect mentioned in the 
introduction. In the following, let us provide a simple intuitive account for the physical origin of 
the wave packet pulsation that we have just seen, in the viewpoint of the dressed state 
picture\cite{dressed state}. 

\subsection{Dressed state picture, on-resonance}

Dressed states are the eigenstates of the atom-cavity combined system which form a manifold of infinite 
hierarchy called the Jaynes-Cummings ladder. Since only one quantum is in the system, we need only to 
consider up to the first couplet such that
\be
|\pm\rangle=(1/\sqrt2)(|0,e\rangle\pm i|1,g\rangle)
\ee
which have the Rabi-split eigenenergies given by
\be
 \label{Epm}
 E_{\pm}(x)=\hbar[\omega_0\pm g_0\sin(kx)]
\ee 
where 0 and 1 count the number of quanta in the field. FIG.\,\ref{dressed-E}(a) shows the spatial 
variation of $E_\pm(x)$ on the cavity axis around a node. In the figure, one can immediately see that 
if the system is in state $|+\rangle$, it will start to slide down the slope of its energy curve and 
move to the left, and in $|-\rangle$, to the right. However, as one can easily show, the state function 
of the system at an arbitrary time turns out to be always a 50-50 superposition of $|+\rangle$ and 
$|-\rangle$ when the system starts out in an energy eigenstate, e.g., $|1,g\rangle$. Thus, half of the 
wave packet sitting on a node is pushed to the left and the other half, to the right, resulting in the 
splitting of the packet into two pieces. When the two split bumps of wave packets hit the turning points 
at the neighboring nodes, they return just like a pair of classical particles. This is the essential 
explanation as to why we have such pulsation of wave packet as in FIGs.\,\ref{Px-Fock}(a) and (b). We 
can do a very simple Newtonian mechanics here again. One of the two components of the wave packet will 
receive from the energy slope a force of size $F=\hbar k g_0\cos kx$. In the early stage of time where 
linear approximation is valid, $F\sim\hbar k g_0$ and the peak will gain the momentum growing as 
$\bar p\sim\hbar k\tau$ initially. But the overall behavior is rather close to our eye-estimation 
of $\bar p(\tau)\sim0.9\hbar k\tau$ in Eq.\,(\ref{x-estimation}). One can easily show that the period 
$\tau_0$ of the wave packet pulsation in the coordinate space is exactly given by $\tau_0={2\over\sqrt
\mu}\int_0^{\pi/2}{d s\over\sqrt{\sin s}}\approx400$. Cf.\ FIG.\,\ref{Px-Fock}(b).

In this fashion, the dressed-energy gradient created by the nonuniform spatial structure of the cavity 
field plays as a dressed-state selector, splitting the atomic wave packet into two distinct groups of 
orthogonal dressed states, $|+\rangle$ and $|-\rangle$, which is the very OSGE. Let us add that, on an 
antinode, on the other hand, the $|+\rangle$ component of the wave packet tends to be pulled out to 
both sides of the antinode while the $|-\rangle$ component tends to be squashed toward the antinode as 
illustrated in FIG.\,\ref{dressed-E}(b), giving the net effect of distorting the initial Gaussian packet 
into a wing-broadened, peak-sharpened feature. 

\subsection{Off-resonant case}

In the actual experiments, however, it is not practically possible to keep the {\it exact} resonance 
indefinitely. So let us consider the case in which the atom and cavity are slightly detuned. Let
$\delta\equiv\omega_C-\omega_A$ denotes the detuning, where $\omega_{A(C)}$ is the atomic 
transition(cavity resonance) frequency. The atom-cavity detuning causes anti-crossing of the energy 
levels which makes the picture slightly complicated. The first couplet is now designated 
by 
\bm
\bea
 |d_+\rangle&=&\cos\theta|0,e\rangle+i\sin\theta|1,g\rangle\\
 |d_-\rangle&=&\sin\theta|0,e\rangle-i\cos\theta|1,g\rangle
\eea
\em
where 
\be
\label{tan_2t}
 \tan\theta=[\delta/2g(x)]+\sqrt{[\delta/2g(x)]^2+1} 
\ee
with the associated eigenenergies given by
\be
E_1^{\pm}(x)=\hbar[\omega_0+(\delta/2)]\pm\hbar\sqrt{g^2(x)+(\delta/2)^2}
\ee
as depicted in FIG.\,\ref{dressed-detuned}\cite{dressed state}. One has to be careful about association 
of the eigenenergies to the eigenstates. In the region where $X\equiv\delta/2g(x)$ is positive, $|d_\pm
\rangle$ are associated with $E_1^{\pm}$, but in the negative $X$ region, these are associated with 
$E_1^{\mp}$. This is simply because $\sin\theta|_{\pm X}=\cos\theta|_{\mp X}$ (due to the 
mutually-symmetric behavior of $\sin\theta$ and $\cos\theta$ around $\theta=\pi/4$), and the 
eigenenergies are determined by the mod-squares of the coefficients, i.e., 
\bea
E_+&=&\left|\cos\theta\right|^2\hbar\omega_a+\left|\sin\theta\right|^2\hbar\omega_c\;\;\;\mbox{for }X\ge0
\nonum
   &=&\left|\sin\theta\right|^2\hbar\omega_a+\left|\cos\theta\right|^2\hbar\omega_c\;\;\;\mbox{for }X<0,
\eea
etc. It is sufficient to consider the solution of Eq.\,(\ref{tan_2t}) in the range $0\le2\theta\le\pi$. 
Let us first consider two limiting cases. Firstly, in the regions relatively far from the node where 
$|X|\ll1$, we have $\theta\rightarrow\pi/4$, and therefore $|d_\pm\rangle$ respectively approximate to 
$|\pm\rangle$. Thus in these regions, the wave packet behaves just like in the case of exact-resonance 
discussed above, as expected. Note that, in the left hand side of the node, $|d_+\rangle$ has the lower 
energy $E_1^-(x)$ which approaches $E_+(x)$ in Eq.\,(\ref{Epm}); and $|d_-\rangle$, the higher energy 
$E_1^+(x)\approx E_-(x)$, whereas in the right hand side, the levels are reversed. Secondly, in the region 
near the node on the other hand, we have $\delta/2g(x)\rightarrow\pm\infty$. When $g(x)$ approaches the 
node from the positive side, $\theta\rightarrow\pi/2$, and when it approaches from the negative side, 
$\theta\rightarrow0$. Thus the behavior of the first couplet is summarized as
\bea
 \fbox{$g(x)<0$}\;\;\; & \; \fbox{$g(x)>0$}\nonum
 \matrix{E_1^+:\hspace{5.2mm} |d_-\rangle\rightarrow-|1,g\rangle\cr 
 \hspace{-1mm}E_1^-:\hspace{5.2mm}|d_+\rangle\longrightarrow|0,e\rangle}
 &\ \vrule \ \matrix{|d_+\rangle\longrightarrow|1,g\rangle\cr 
 |d_-\rangle\longrightarrow|0,e\rangle}
\eea
What it means is the following: if the system is in $|1,g\rangle$, for instance, the system is placed in 
a potential well defined by $E_1^+(x)$ in this region, and when it is in $|0,e\rangle$, it is sitting on 
a potential hill defined by $E_1^-(x)$. Thus if the system starts from $|1,g\rangle$, the portion of the 
wave packet in this region tends to be trapped in the potential well of $E_1^+(x)$ while the wings of the 
packet outside this central region will slide down the slopes of $E_1^-(x)$ on both sides of the node, 
since outside the region, $|1,g\rangle$ is a combination of $|d_\pm\rangle\approx|\pm\rangle$ as 
aforementioned. Note that the overall state of the system at arbitrary time is no longer a 50-50 
combination of $|d_\pm\rangle$. Nevertheless, the packet motion occurs always in a symmetric fashion about 
the node, as a matter of course. To see this, write the internal state as $|\psi\rangle=C_1|0,e\rangle
+C_2|1,g\rangle$ with complex coefficients $C_1$ and $C_2$. Then, at a position $x=x_1$ (with respect to 
the origin at a node), it is written as $|\psi\rangle=A_+|d_+\rangle+A_-|d_-\rangle$ where $A_+=C_1\cos
\theta-iC_2\sin\theta$ and $A_-=C_1\sin\theta+iC_2\cos\theta$, whereas at $x=-x_1$, it is given by $|\psi
\rangle=B_+|d_+\rangle+B_-|d_-\rangle$ where $B_+=C_1\sin\theta-iC_2\cos\theta$ and $B_-=C_1\cos\theta
+iC_2\sin\theta$. But since $|A_\pm|=|B_\mp|$, the portions of the wave packet in the upper energy state 
on both sides are the same. Thus the symmetry arises around the node. The ratio of the portions in the 
upper and the lower energy states, however, are changing in time. As a result, we will have the initial 
single-bumped wave packet evolving into a {\it triple}-peaked structure on a node in the presence of 
finite atom-cavity detuning instead of the double-peaked shape in case of the exact resonance. 

Our expectation is nicely confirmed by a numerical calculation as shown in FIG.\,\ref{Fock-detuned} where 
we take $\delta/g_0=0.1$, for instance, with other parameters the same as before. Of course, due to the 
small effective coupling constant, the system slowly evolves from $|1,g\rangle$ to $|0,e\rangle$, and the 
portion of the wave packet in the central region will also gradually migrates into both sides of the node. 
The volume of the central region is confined by $|(2g_0/\delta)\sin(kx)|\ll1$ or $|x|/\lambda\ll(1/4\pi)
(\delta/g_0)$. For $\Delta x/\lambda=1/40\pi$, we have $|x|/\Delta x\ll10(\delta/g_0)$. If $\delta/g_0
\sim0.01$, then $|x|/\Delta x\ll0.1$, which implies, though loosely, that by far the most part of the wave 
packet resides outside this region and the central peak will not be significant if $\delta/g_0\le0.01$. 

Let us leave a brief note regarding the dependence on the field state. If the field state includes a nonzero
amplitude of the vacuum state $|0\rangle$ as in the case of a coherent state, this portion of field will 
tend to leave the atomic wave packet intact in time. So, there will be always some portion of the packet 
standing still at its original position. This will show up as a peak at the center between the two separate 
peaks on both sides, making the overall look of the wave packet somewhat similar to the case of nonzero 
detuning aforementioned. This fact is most easily seen in the Schr\"odinger picture. We will get back to 
this point at an appropriate place in the next section since we will explicitly deal with the Schr\"odinger 
equations there. 

\section{The effect of damping}
\label{sect3}

The picture given so far in the dressed-state formalism provided a qualitative but intuitive understanding 
of the phenomenon. Nevertheless, the real systems are after all {\it open} quantum systems subject to 
damping, and the dynamical effects of such decoherent processes are yet to be discussed. The open systems 
can be dealt with by a number of ways, e.g., the master equations\cite{qtt,gardiner}, the quantum Langevin 
equations\cite{gardiner}, or the Fokker-Planck equations\cite{gardiner,carm2} in appropriate situations. 
However, since we are dealing with the dynamics of a {\it single} atomic wave packet instead of a statistical 
ensemble, we find it most appropriate to look into the system in the viewpoint of the quantum trajectory 
theory (QTT)\cite{qtt}. The theory provides most direct and intuitive descriptions as to the dynamics of such 
open quantum systems. Furthermore, as we will show below, the QTT suggests an interesting phenomenon that the 
atomic spontaneous emissions in fact catastrophically split the wave packet into two or more peaks when the 
atom is located at or moving through a node of the field.  

A quantum trajectory consists of the piece-wise continuous coherent evolution and the discontinuous quantum 
jumps, such as the atomic and cavity decays, which randomly interrupt the coherent evolution. Let $\Gamma_i$ 
$(i=1,2,...)$ denote the various decay rates in the system, and $C_i$ the corresponding ``collapse'' operators, 
e.g., $\sigma_-$ for the atomic emission and $a$ for the cavity transmission. Then the coherent evolution of the 
system is given by the following Schr\"odinger-like equation
\be
 \label{coherent process}
 i\hbar{\d\over\d\tau}|\bar{\Psi}\rangle=\left(H-\sum_ii\hbar{\Gamma_i\over2}C_i^\dag C_i\right)
|\bar{\Psi}\rangle\,.
\ee
Thus the Schr\"odinger process is slightly modified by the damping terms. Because of the damping terms, it 
is not a norm-conserving process, and $|\bar{\Psi}\rangle$ represents an {\it unnormalized} wave function.
When the collapse probability 
\be
 p_c=\Gamma\dt{\langle\bar\Psi|C^\dag C|\bar\Psi\rangle\over\langle\bar\Psi|\bar\Psi\rangle}
\ee
is greater than a random number $R\in[0,1)$ taken during the time interval $[t, t+\dt)$, a quantum jump 
occurs in the fashion
\be
|\bar\Psi\rangle\longrightarrow C|\bar\Psi\rangle\,.
\ee
Otherwise, the system dynamics follow Eq.\,(\ref{coherent process}). For further details of the theory, 
see, e.g., Ref.\cite{qtt}.

Now let us return to the original problem in which an atomic wave packet in its ground state is placed at a 
node of the cavity field mode, on the axis of the cavity. The cavity can contain any photonic state of light, 
and let us neglect the cavity damping ($\kappa=0$) for the time being in order to focus on the effect of the 
atomic damping, and assume a very massive atom so that the kinetic energy term can be neglected (Raman-Nath 
approximation). For algebraic convenience, let us write $g(x)=g_0\cos(kx)$, and assume a very massive atom 
so that the kinetic energy term can be neglected (Raman-Nath approximation). Then the Hamiltonian in the 
interaction picture becomes as simple as
\be
\label{hamiltonian1}
H=i\hbar g_0\cos(kx)\left(\sigma_-a^\dag-\sigma_+a\right)\,.
\ee
We expand the wave function in the fashion,
\wide
\be
\label{psi-function}
 |\bar{\Psi}\rangle=\int \d q\left\{\sum_{n=1}^\infty\left[\bar{C}_e(n,q)|n-1,q,e\rangle
+\bar{C}_g(n,q)|n,q,g\rangle\right]+\bar{C}_g(0,q)|0,q,g\rangle\right\}
\ee
\narrow
where $n$ is the field quantum number and $q$ the atomic momentum quantum number scaled in unit of $\hbar k$. 
The coherent evolution is then given by the following set of dynamical equations:
\wide
\bm
\bea
 \label{d-eq1}
 {\d\over\d\tau}\bar{C}_e(n,q)&=&-{\sqrt n\over2}\left[\bar{C}_g(n,q-1)+\bar{C}_g(n,q+1)\right]
-{\gamma\over2}\bar{C}_e(n,q)\hspace{.5cm}(n\ge1)\\
 \label{d-eq2}
 {\d\over\d\tau}\bar{C}_g(n,q)&=&+{\sqrt n\over2}\left[\bar{C}_e(n,q-1)+\bar{C}_e(n,q+1)\right]
\hspace{2.43cm}(n\ge0)
\eea
\em
\narrow
where we used the relations $\cos(kx)=(1/2)[\exp(ikx)+\exp(-ikx)]$ and $\exp(\pm ikx)|q\rangle=|q\pm
1\rangle$ with $\tau=g_0t$, and $\gamma=\gamma_A/g_0$, the atomic decay rate scaled in $g_0$. So the 
$n$-th Jaynes-Cummings couplet is decoupled in dynamics from the rest of the infinite hierarchy of 
couplets. Furthermore, it is trivially seen that the vacuum state component $C_g(0,q)$ does not change 
in time. 

The state of the cavity field $|\psi_{\rm f}\rangle$ can be always expressed as a linear combination 
of number states, i.e.,
\be
 |\psi_{\rm f}\rangle=\sum_nf(n)|n\rangle\,.
\ee
Let the initial atomic position $x=\xi_0\lambda$ and the momentum spread $\Delta p=(\Delta q)\hbar k$. 
Then the initial wave packet is given by 
\wide
\be
 \label{wf-init1-q}
 |\Psi(0)\rangle=\int\d q\sum_{n=0}^\infty\exp\left[-q^2\over4(\Delta q)^2\right]
\exp\left(-2\pi iq\xi_0\right)F(n)|n,q,g\rangle
\ee
\narrow
where $F(n)$ denotes the normalized coefficients such that
\be
 F(n)={1\over\root4\of{2\pi(\Delta q)^2}}f(n)\,.
\ee
The initial values are therefore
\bm
\bea
 \label{node-init-cf1}
 C^0_e(n,q)&=&0\hspace{.5cm}(n\ge1)\\
 \label{node-init-cf2}
 C^0_g(n,q)&=&F(n)\exp\left[-q^2\over4(\Delta q)^2\right]\exp\left(-2\pi iq\xi_0\right)\,.
\eea
\em
Now at a node, $\xi_0=1/4$, and Eq.\,(\ref{node-init-cf2}) is written as
\bea
 C^0_g(n,q)=F(n)e^{-\beta q^2}e^{-i{\pi\over2}q}
\eea
where $\beta=1/4(\Delta q)^2$, being the width-parameter of the initial wave packet. With this initial 
value, we solve Eqs.\,(\ref{d-eq1}) and (\ref{d-eq2}) in the Eulerian fashion such that
\be
\bar C^{\nu\d\tau}_{e(g)}(n,q)=\bar C^{(\nu-1)\d\tau}_{e(g)}(n,q)-\d\tau\dot{\bar C}^{(\nu-1)
\d\tau}_{e(g)}(n,q)
\ee
where time $\tau$ is divided by small discrete steps $\d\tau$ so that $\tau=\nu\d\tau$ with an 
integer $\nu$. We obtain the behavior of the system at an early time $\tau$ given by
\wide
\bm
\bea
 \label{D_e^t}
 \bar{C}^{\tau}_e(n,q)&\sim&-iC_g^0(n,q)\left(\tau\sqrt n\over2\right)\left[e^{\beta(2q-1)}
-e^{-\beta(2q+1)}\right]e^{-{\gamma\over2}\tau}\hspace{1.5cm}(n\ge1)\\
 \label{D_g^t}
 \bar{C}^{\tau}_g(n,q)&\sim&C_g^0(n,q)\left\{1+{3\over8}\left(\tau\sqrt n\over2\right)^2
\left[e^{4\beta(q-1)}-2+e^{-4\beta(q+1)}\right]\right\}\hspace{.45cm}(n\ge0)
\eea
\em
\narrow
keeping terms up to the order of $(\d\tau)^2$. Note that by comparing with the numerical solutions, 
we find these expressions are valid up to $\tau\sim5$. [At this moment, let us take a brief pause and 
look at the dependence of the shape of the split wave packet on the field state mentioned in the 
last section. The momentum distribution ${\cal P}(q)$ at an arbitrary time is given by
\bea
 {\cal P}(q)&=&\sum_{n=1}^\infty\left[\left|C^\tau_e(n,q)\right|^2+\left|C^\tau_g(n,q)\right|^2\right]
+\left|C^0_g(0,q)\right|^2\nonum
 &=&\cdots+{|f(0)|^2\over\root\of{2\pi(\Delta q)^2}}\exp\left[-{q^2\over2(\Delta q)^2}\right]\,.
\eea
So it is clearly seen that when the field state is contaminated by a nonzero vacuum state component, 
a miniature of the original Gaussian packet will always be seen between the separated wings.] If the 
damping is weak, there will be little difference in the coherent evolution in the early times during 
which $\exp(-\gamma\tau/2)\approx1$. {\it But what it does in the jump process is remarkable}. Since 
now there is a small probability built up that the atom is in the excited state, there is a 
probability that the atom can decay. Let us just assume that such a decay happened now. The wave packet 
then undergoes a quantum jump such that
\be
\label{atom decay}
 |\bar\Psi\rangle\longrightarrow\exp(-ikx\eta)\sigma_-|\Psi\rangle
\ee
where $\exp(-ikx\eta)$ describes the momentum recoil that the atom gets from the emission, projected 
on the $x$-axis, and $\eta$ is a random number in the range $[-1,1]$. $|\bar\Psi\rangle$ represents an 
unnormalized wave function. In the coefficients, this process is written as
\wide
\bm
\bea
  \bar C^\tau_e(n,q)&\longrightarrow&0\\
  \bar C^\tau_g(n,q)&\longrightarrow&\bar C^\tau_e(n+1,q+\eta)\propto F(n+1)
   \left[e^{-\beta(q+\eta-1)^2}-e^{-\beta (q+\eta+1)^2}\right]e^{-{\gamma\over2}\tau}.
\eea
\em
\narrow
Since then, the entire wave packet is determined by $\bar C^\tau_g(n,q)$, yielding the probability 
distribution in the momentum space such that 
\bea
 \label{Pg-simple}
 {\cal P}(q)&=&\sum_n\left|C^\tau_g(n,q)\right|^2\nonum
 &\propto&\left[e^{-\beta(q+\eta+1)^2}-e^{-\beta(q+\eta-1)^2}\right]^2\,.
\eea
It is simply the square of the difference of two Gaussians which are slightly shifted from each other. 
So obviously it will show a {\it double}-peaked structure. The locations of the extrema of ${\cal P}(q)$ 
are easily found in the limit of small $\beta$. Note that we are considering a spatially well-localized 
atomic wave packet $(\Delta x\ll\lambda)$ to which the small $\beta$ limit applies. Then the locations 
of the two bumps in ${\cal P}(q)$ are given by $q\approx\pm\sqrt2(\Delta q)$ $(|\eta|\ll\Delta q)$. Thus 
the initial momentum spread somehow determines the locations of the peaks in the split wave packet in 
the early stage of atom-field interaction. 

FIGs.\,\ref{jumps-2D}(a) and (b) show the shapes of the wave packet sitting on a node right after an 
atomic jump arranged to occur at a few different times in the early stage of atom-field interaction: (a) 
in the momentum space and (b) in the coordinate space, at various times of the atomic jump, $\tau=1$, 2, 
$\cdots$, 5, 10 and 20 with the same set of parameters as in FIG.\,\ref{typical-jump}. The Gaussian 
curve is the initial packet. Note that the curves are independent trajectories. The figures show that 
our simple argument reaching Eq.\,(\ref{Pg-simple}) is quite valid up to $\tau\sim5$. It is seen that 
in the coordinate space, the longer the coherent evolution time elapsed before the jump, the narrower 
both the widths of, and the separation between, the split peaks become, whereas in the momentum space, 
the wider both become. 

In fact, one can extract two separate dynamical mechanisms at work in these pictures. Firstly, the 
packet splitting in the coordinate space---or in other words, the fact that ${\cal P(\xi)}=0$ at a 
node---right after an atomic jump has the following physical ground. The event of an atomic jump 
implies that the atom was in the excited state before the jump. For the atom to be lifted from its 
initial ground state to the excited state, the atom-field interaction must be nonzero. So an event of 
the atomic jump tells us that the probability for the atom to be at the node was indeed zero, and our 
initial Gaussian wave packet has been accordingly modified through the back-action of the measurement. 
Secondly, the reason that the separation of the two split peaks in the coordinate space becomes smaller 
and smaller as the advent of the atomic jump is delayed is explained in the same context. That is, if 
the atomic jump has not occurred after all, we have to conclude that actually the atom has not gone 
through the Rabi interaction with the cavity field, and has stayed in the ground state. This means that 
the wave packet becomes narrower and narrower around a node as time goes by before an atomic jump 
really occurs. When a jump occurs, the packet is split into two through the first mechanism, but since 
the separation of the split packet is determined by the width of the packet before the jump, it is 
narrower also. Thus one may add that the theory of the quantum trajectory is none other than the 
process of continuous correction of our ``lack of knowledge'' about the state of the system, based upon 
the information provided from the measurements. The measurements here are of course made by the 
environment---i.e., the vacuum field which is continuously monitoring the atomic damping.

The irregular shifts of the packets in the momentum space are due to the random momentum recoils from 
the atomic spontaneous emissions, whereas no such shifts are shown in the coordinate space. Note, 
however, that the latter is {\it not} because the atomic mass is infinite, but simply because 
\bea
|\psi_1(x)|^2&=&\left|\int\d p\varphi(p-p_1)\exp(ipx/\hbar)\right|^2\nonum
           &=&\left|\int\d p\varphi(p)\exp(ipx/\hbar)\right|^2=|\psi_0(x)|^2
\eea
regardless of the size of the recoil momentum $p_1$ or the atomic mass, for the dual wave functions 
$\psi(x)$ and $\varphi(p)$. The physical reason of this is that the atomic jump has been assumed to 
occur {\it instantaneously}, i.e., no time has elapsed before and after the jump. So there is no 
displacement change before and after the momentum kick. However, the absence of motion in the coordinate 
space between the atomic jumps is due to the infinity of the atomic mass. As a matter of course, when the 
atomic mass is finite, the system will evolve differently in the coordinate space {\it after} an atomic 
jump for a different size of the momentum kick that it has gained. Anyhow, we see that there is some 
``internal-motion,'' if we may, i.e., splitting, spreading and separating, of the wave packet even for an 
infinitely massive atom in an optical cavity although there is no center-of-mass motion. This interesting 
feature seems to deserve some further discussion. The following subsection is devoted to this purpose. 

\subsection{The ``dynamics'' of an infinitely heavy atom.}

Notice that this wave packet splitting {\it in the coordinate space} by the atomic decays have occurred 
even for an infinitely heavy atom. Normally, in the usual concept of the wave packet ``dispersion'' 
introduced in the textbooks, the mass of the particle matters. That is, if an infinitely heavy particle 
is in a free space, its wave packet will show no dynamics---no center-of-mass motion or any distortion 
of its shape at all, either in the momentum or in the coordinate space---as a matter of principle. For 
the present cavity-QED system with an infinitely heavy atom, however, some interesting features appear. 
The atomic wave packet of an infinite mass does show dynamics in the momentum space, i.e., splitting, 
spreading and separating, which is of course due to the momentum exchanges with the field. But in the 
coordinate space, nothing happens provided that the system is free of damping, {\it no matter what 
happens to the wave packet in the momentum space}, as shown in FIGs.\,\ref{no-damp}(a) and (b). In the 
figure, the inset is a lateral view of the 3D-plot on the bottom from the time axis which clearly shows 
the time-invariance of ${\cal P}(\xi)$, the atomic wave packet in the coordinate space. In the 
mechanical point of view, this seems rather understandable considering that the atomic mass is assumed 
to be infinite, but is still quite interesting in the sense that ${\cal P}(\xi)$ is essentially (but not 
exactly in fact) a Fourier-type transform of ${\cal P}(q)$. So, if ${\cal P}(q)$ changes so radically as 
shown in FIG.\,\ref{no-damp}(a), one would na\"\i vely expect some change in ${\cal P}(\xi)$ also, 
whatever it may be. 

If we write the wave packet in the same fashion as Eq.\,(\ref{psi-function}), ${\cal P}(\xi)$ at time $t$ 
is given by
\be
{\cal P}(\xi)={\cal P}_0\sum_{n,a}\left|\int\d qe^{i2\pi q\xi}C_a(n,q)\right|^2
\ee
where $a$ labels the atomic internal states and ${\cal P}_0$ a normalizing constant. The coefficients $C_a(n,q)$ 
continuously evolve in time with the Hamiltonian, Eq.\,(\ref{hamiltonian1}), but only in such a fashion that 
${\cal P}(\xi)$ remains invariant in time. This is in fact easily proved by looking at the time derivative of 
${\cal P}(\xi)$, i.e.,
\wide
\bea
{\d\over\dt}{\cal P}(\xi)&=&{\cal P}_0\sum_{n,a}\int\d q\d q^\prime e^{i2\pi(q-q^\prime)\xi}
\left[\dot C_a(n,q)C_a^\ast(n,q^\prime)+C_a(n,q)\dot C_a^\ast(n,q^\prime)\right]\,.
\eea
\narrow
Now from the dynamical equations, Eqs.\,(\ref{d-eq1}) and (\ref{d-eq2}), without the damping term, we have
\wide
\bm
\bea
 \label{d-eq11}
 \dot{C}_e(n,q){C}^\ast_e(n,q^\prime)&=&-{\sqrt n\over2}\left[{C}_g(n,q-1){C}^\ast_e(n,q^\prime)
+{C}_g(n,q+1){C}^\ast_e(n,q^\prime)\right]\,,\\
 \label{d-eq222}
 {C}_g(n,q)\dot{C}^\ast_g(n,q^\prime)&=&+{\sqrt n\over2}\left[{C}_g(n,q){C}^\ast_e(n,q^\prime-1)
+{C}_g(n,q){C}^\ast_e(n,q^\prime+1)\right]\,,
\eea
\em
etc., and simply because 
\be
 \label{funny-sum}
 \int\d q\d q^\prime e^{i2\pi(q-q^\prime)\xi}\left[-{C}_g(n,q-1){C}^\ast_e(n,q^\prime)
+{C}_g(n,q){C}^\ast_e(n,q^\prime+1)\right]=0\,,
\ee
\narrow
etc., we find
\be
{\d\over\dt}{\cal P}(\xi)=0\,.
\ee

Nevertheless, even more interesting things happen when the atom is subject to damping. Then the wave packet 
of an infinitely massive atom does show dynamical evolutions in the coordinate space as well via the coherent 
decay during the continuous evolution and by the atomic jumps. FIGs.\,\ref{typical-jump}(a) and (b) show a 
typical time evolution of a wave packet which has gone through a single atomic jump in the momentum space (a) 
and in the coordinate space (b), where $\gamma=0.1$, $\Delta q=10$, and $M=\infty$ for an atom initially in 
the ground state and the field in a one-quantum coherent state in a lossless cavity. 
In FIG.\,\ref{typical-jump}(b), we clearly see the actual evolution of the wave packet in the coordinate 
space during the coherent process, not only the drastic change after the atomic jump: The wave packet gets 
indeed continuously squeezed around its peak. The question is how such a motion of the wave packet of an 
infinitely heavy atom can occur in the coordinate space. The reason that damping causes this type of wave 
packet evolution even for an infinitely massive atom can be seen in Eqs.\,(\ref{d-eq1}) and (\ref{d-eq2}). 
There the atomic damping makes the two equations asymmetric in the sense that it adds a damping term only in 
Eq.\,(\ref{d-eq1}), and thereby we no longer have the identities such as Eq.\,(\ref{funny-sum}). 

Since everything happened for an infinitively heavy atom, the dynamics following an atomic jump is {\it not} 
a mechanical process. That is, the mean position of a peak in the coordinate space cannot be related to a 
time integral of that in the momentum space in the fashion $\bar x=\int(\bar p/M)\dt$, as we did in 
Sect\,\ref{sect2}, for an obvious reason. The alert readers will find that the correct relation between the 
mean positions $\bar x$ and $\bar p$ in both spaces right after an atomic jump is indeed obtained from the 
uncertainty relation, i.e., $\bar x\sim\alpha/\bar p$ where $\alpha$ is a constant on the order of $\hbar$, 
though it is not exactly $\hbar/2$ because the wave packet no longer retains the minimum uncertainty, as 
FIGs.\,\ref{typical-jump}(a) and (b) demonstrate. Thus, it should be entirely attributed to the intrinsic 
wave nature of a matter particle as well as to the issue of the position and momentum uncertainty which 
forms the very heart of the quantum mechanics\cite{bohm}. 

\subsection{Over-damped case}

When the atomic damping is strong such that $\gamma\ge1$ (out of the strong coupling regime), the wave packet 
essentially follows the coherent decay only, without quantum jumps. This can be quickly appreciated from the 
dynamical equations in the quantum trajectory, Eqs.\,(\ref{D_e^t}) and (\ref{D_g^t}). The form of 
Eq.\,(\ref{D_e^t}) shows that the atomic upper-state probability $|C_e(n,q)|^2$ has an early time dependence 
roughly of the form $\tau^2\exp(-\gamma\tau)$. FIG.\,\ref{Pe} shows the temporal behaviors of ${\cal P}_e(q)=
\sum_n|C_e(n,q)|^2$ and $P_e=\int\d q{\cal P}(q)$ in a trajectory in which an atomic jump has not occurred yet. 
For a large $\gamma$, the exponential-type factor dictates so that the atomic upper-state probability quickly 
decays to zero, having no time to grow up to an appreciable value. Thus no atomic jumps are likely, and the 
system follows only the coherent decay. While damping tends to put the system back to its initial internal 
energy state in this fashion, the atom-field momentum exchange process rapidly broadens the wave packet in the 
momentum space, which ends up narrowing---i.e., localizing---the wave packet in the coordinate space although 
the uncertainty product $\Delta x\Delta p$ slowly grows in general. This is none other than the phenomenon of 
the ``damping-induced localization'' (DIL)\cite{zurek}. For a moderate atomic damping, the wave packet follows 
the coherent evolution showing the DIL until an atomic jump which brings about a sudden splitting---a gigantic 
change---in the wave packet as shown in FIGs.\,\ref{typical-jump}(a) and (b). 

\subsection{Successive atomic jumps}

The effect of successive atomic decays are studied through a numerical calculation, the results of which are 
displayed in FIGs.\,\ref{regular-jumps} (a) and (b), we set $\gamma=0.5$, $\kappa=0$ and $\Delta q=10$ and 
$M=\infty$ with the initial field in a one-quantum coherent state $|\alpha=1\rangle$. To quickly appreciate 
the dynamics, we let the system undergo spontaneous decays at $\tau=5,10,15$, etc.  It is shown that the first 
atomic decay splits the momentum space wave packet into two peaks, the second into three, the third into four, 
and so on. However, the coordinate space wave packet remains largely as a double-peaked shape ever since the 
first atomic decay as shown in FIG.\,\ref{regular-jumps} (b).  

\subsection{Effect of cavity decay}

The cavity quantum jumps, on the other hand, do not do much in the wave packet splitting. To see this let us 
turn off the atomic damping channel and open the cavity decay channel. Then Eq.\,(\ref{atom decay}) is 
replaced by the process
\be
\label{cavity decay}
 |\bar\Psi\rangle\longrightarrow a|\Psi\rangle
\ee
which brings about the changes in the coefficients such that
\bm
\bea
  \bar C^\tau_e(n,q)&\longrightarrow&\sqrt{n}C^\tau_e(n+1,q)\\
  \bar C^\tau_g(n,q)&\longrightarrow&\sqrt{n+1}C^\tau_g(n+1,q).
\eea
\em
It will certainly change the shape of the wave packet slightly, but never as much as splitting the packet into 
two pieces. Particularly when the mean cavity photon number is high, there will be no essential change, whereas 
the packet splitting due to the atomic decay is regardless of the intensity of the field. 

\section{Experimental consideration}
\label{sect4}
Since the atomic damping changes the wave packet in such a radical fashion, one may expect that it can play a 
significant role in the actual experiments. It is therefore meaningful to compare the result of a hypothetical 
experiment in which the atoms assumed to be nonradiative with that of a realistic experiment in which the atoms 
are radiative. The actual experimental setup will be most possibly such that a beam of atoms are launched to 
fly through the cavity mode while the cavity is continuously pumped by an external field, and one measures the 
position distribution of the atoms emerging from the interaction region on a surface at some distance from the 
cavity. Since there will be always some fraction of atoms which will pass nonradiatively through the interaction 
region without undergoing any jumps even in the presence of nonzero damping strength, one can expect some nonzero 
portion of the undeflected atoms in the resulting pattern of the position distribution. However, if one can still 
observe two or more split bumps in the background, one might say, it is those atoms which have undergone any 
number of jumps that are responsible for the observed wave packet splitting, particularly considering the fact 
that the wave packet splitting through the coherent process tends to be suppressed (by the DIL aforementioned, 
which will be seen in FIGs.\,\ref{fig-exp} and \ref{fig-exp-no-jump} shortly) in the presence of appreciable 
damping.

Since it is difficult to analytically discuss the outcome of such an experiment, we have performed Monte Carlo 
simulations based on the quantum trajectory theory. As we did in our previous work on the cavity-QED atom 
detection system\cite{tak}, we assume a single mode cavity resonant to the transition line of barium ($^{138}$Ba) 
atom, $\lambda=553$\,nm, with $g_0/2\pi=42$\,MHz, which correspond to $\mu\sim1.1\times10^{-4}$. The transverse 
cavity mode is assumed to have a profile so that $g(\vec x)=g_0\exp[-(y^2+z^2)/w^2]\cos kx$, where the cavity 
axis lies in the $x$-direction, and $w$, the mode waist is taken to be $37$\,$\mu$m. The atomic longitudinal 
velocity is set $v_z=400$\,m/sec. The driving field amplitude $\cal E$ is such that ${\cal E}/\kappa=1$ so that 
the mean intracavity photon number is just one at steady state, i.e., $\langle a^\dag a\rangle_{ss}=1$, before 
the entry of an atom. We choose the damping strengths as large as $\gamma/g_0=2\kappa/g_0=0.5$. We assume that 
the initial atomic wave packet, to be determined by the geometry of the system, is a Gaussian having the momentum 
uncertainty $\Delta p=10\hbar k$ which again corresponds to $\Delta x\le\lambda/100$, a well-localized, 
particle-like wave packet. Note that the ``initial time'' is the time when the atoms reach the plane normal to 
the atomic path, assumed to be at $10w$ away, for instance, from the cavity axis. Thus, until the wave packet 
reaches the region of appreciable atom-field coupling strength, it will freely evolve.

FIG.\,\ref{fig-exp} shows the ``far-field'' position distributions averaged over 200 atoms detected on a screen 
located at a distance---approximately $40w$ away, for instance, from the cavity axis. (Note that the far-field 
position distributions are exactly the same in shape as the momentum distributions with proper axis labeling 
and scaling, according to the diffraction theory.) The thinner line is for the case (a) in which the atoms are 
assumed to be nonradiative, while the thicker is for the case (b) where the atomic decay channel is let open with 
$\gamma/g_0=0.5$. It turns out that during the entire flight time, each atom has gone through about 4.4 jumps on 
average in case (b). The figure shows that, even for the atoms and the cavity as strongly damped as the given 
strengths, the wave packet splitting robustly shows up, having hardly been washed out. The overall width of 
${\cal P}(\xi)$ ($\Delta\xi\sim0.68$) in case (a) is greater than the value ($\Delta\xi\sim0.46$) in case (b), 
the height of ${\cal P}(0)$ in case (a) being smaller than that in case (b). The narrower distribution of case 
(b) can be interpreted as weaker diffusive motion of the atoms in the field during the interaction time. Less
diffusion implies less frequent exchanges of quanta between the atom and field, i.e., weaker interaction. So the 
mean field intensity is smaller in the latter case, which again indicates greater portion of the vacuum state in 
the cavity field on average. The stronger contamination of the cavity field by the vacuum state is due to the 
introduction of the additional damping channel, i.e., the atomic damping, to the system. The narrower width in 
case (b) can be understood in this way which may be another interpretation of DIL. We suggest that the effect of 
the atomic damping will show up as these differences in the shape of the position distribution of the detected 
atoms. Most notably, the width of ${\cal P}(\xi)$ will be generally narrower than that predicted for the ideal 
systems without damping. 

Let us finally address that the resultant wave packet splitting can be entirely attributed to the discontinuous 
atomic {\it jumps}, in the viewpoint of the quantum trajectory theory. For this, we consider a fictitious 
situation (c) where we artificially suppressed the atomic jumps, allowing only the continuous decay. The result 
is shown in FIG.\,\ref{fig-exp-no-jump} by the thin single-peaked line which is in comparison with the lower 
heavy one, the same curve for the realistic case (b) shown in FIG.\,\ref{fig-exp}. As seen in the figure, the 
coherent process alone did not cause splitting of the wave packet. On the contrary, it has actually prohibited 
the spreading of the wave packet in the coordinate space. (The wave packet would have been much wider than both 
curves even in the free space at the same elapsed time.) So it appears that what the atomic jumps did is simply 
to curtail the central portion of the distribution in (c) and relocate it to both sides, yielding the 
double-peaked shape in (b). In this regard, we conclude that the wave packet splitting in the case of an 
appreciably strong atomic damping is essentially due to the discontinuous process---in the viewpoint of the 
quantum trajectory theory---rather than the coherent process which the OSGE was originally envisaged to be. Note 
also that, right after an atomic jump, the atomic state is put into the ground state which is simply a 50-50 
mixture of the orthogonal dressed states. Furthermore, in the presence of damping, the final state of the system
will eventually decay into the product of the atomic ground state and the initial coherent field state such that
\be
 \lim_{t\rightarrow\infty}|\Psi(t)\rangle=|\psi_{\rm f}\rangle\otimes|\psi_{\rm a}\rangle
=|\alpha\rangle\otimes\!\int\!\d p\,C(p)|p,g\rangle
\ee 
where $\alpha={\cal E}/\kappa$, since the atom-field entanglement gets lost as the cavity field is restored to 
the initial state by the driving field whereas the atom eventually decays into the ground state. Thus the atomic 
states on both peaks in the final distribution are again the same. Note that without the atomic damping---as the 
ideal theories usually assume, the outcome of such an experiment would yield the wave packet split into a pair 
of distinct bumps at large, like the thin line in FIG.\,\ref{fig-exp}, and each bump will correspond to the 
aggregate of atoms in either of the orthogonal dressed-states. When the atomic damping is present, however, it 
is difficult to say that the state on one peak is, or has been, orthogonal to that on the other because the 
atomic jumps have been constantly putting the atoms on both peaks in the same ground state.

\section{Summary}
\label{sect5}

We have analyzed the optical Stern-Gerlach effect in the realm of the cavity-QED. We included various damping 
mechanisms in the system, and discussed particularly the effect the atomic damping in the wave packet evolution 
in the framework of the quantum trajectory theory. We have shown that the {\it continuous} atomic decay process 
tends to localize the atomic wave packet, keeping it from spreading so fast as in the case of no atomic damping, 
whereas the {\it discontinuous} atomic jumps tend to split the packet. Proposing a possible scheme of an 
experiment, we pointed out that the phenomenon of wave packet splitting could be experimentally observed even in 
a moderate coupling regime where the atomic damping rate is comparable to the atom-field coupling constant, 
producing discernible differences from the predictions made by ideal theories in which the damping effects are 
disregarded.

\section{acknowledgment}
This work was supported by the Creative Research Initiatives of the Korean Ministry of Science and Technology.

\bfg
%\centerbmp{2.5in}{1in}{model.bmp}
\caption{The model.}
\label{model}
\efg

\bfg
%\centerbmp{3.75in}{3in}{Fock-M100-Pp.bmp}
%\centerbmp{3.75in}{3in}{Fock-M100-Px.bmp}
\caption{The evolution of an atomic wave packet placed on a node of the cavity field. (a) In the 
momentum space, and (b) in the coordinate space. $q=p/\hbar k$, $\xi=x/\lambda$ and $\tau=g_0t$. 
For a typical atom with $\mu=\hbar k^2/2Mg_0\sim1.7\times10^{-4}$. The atom is initially in the 
ground state having a momentum spread $\Delta p=10\hbar k$ while the field is prepared in a single 
quantum Fock state. The inset is a lateral view of (b) into the time axis along with the $\cos(kx)$ 
curve.}
\label{Px-Fock}
\efg

\bfg
%\centerbmp{3.2in}{1.5in}{dressed1.bmp}
%\centerbmp{3.2in}{1.7in}{dressed2.bmp}
\caption{Spatial modulation of the dressed-energies ($\hbar=1$) and motion of the dressed-states 
starting from a node (a) and from an antinode (b). ``{\it A}'' and ``{\it N}'' respectively denote 
antinode and node.}
\label{dressed-E}
\efg

\bfg
%\centerbmp{3.2in}{1.5in}{dressed-detuned.bmp}
\caption{Spatial modulation of the dressed-energies ($\hbar=1$) in the presence of nonzero atom-field 
detuning. $\omega^\prime=\omega_0+\delta/2$ and $g^\prime=\sqrt{g_0^2+(\delta/2)^2}$.}
\label{dressed-detuned}
\efg

\bfg
%\centerbmp{3.25in}{2.5in}{Fock-detune01-Px.bmp}
\caption{Wave packet evolution in the coordinate space in the presence of nonzero atom-field detuning, 
$\delta/g_0=0.1$, while other parameters are the same as in FIG.\,\ref{Px-Fock}. $\xi=x/\lambda$.}
\label{Fock-detuned}
\efg

\bfg
%\centerbmp{2.8in}{2in}{jumps-Pp.bmp}
%\centerbmp{2.8in}{2in}{jumps-Px.bmp}
\caption{The shapes of the wave packet sitting on a node right after an atomic jump arranged to occur 
at $\tau=1$, $\cdots$, 5, 10 and 20. $q=p/\hbar k, \xi=x/\lambda$.}
\label{jumps-2D}
\efg

\bfg
%\centerbmp{2.75in}{1.8in}{no-damp-Pp-3D.bmp}
%\centerbmp{2.55in}{1.8in}{no-damp-Px-3D.bmp}
\caption{The time evolution of the atomic wave packet of an infinitely heavy atom free of damping 
located at a node (a) in the momentum space and (b) in the coordinate space. The insets are the 
ground-level views into the time axis.} 
\label{no-damp}
\efg

\bfg
%\centerbmp{2.7in}{1.8in}{jump-Pp-3D.bmp}
%\centerbmp{2.7in}{1.8in}{jump-Px-3D.bmp}
\caption{The dynamics of a wave packet in the presence of damping in which an atomic jump has occurred, (a) 
in the momentum space and (b) in the coordinate space. $\gamma=0.1$ and $\Delta q=10$, for an {\it infinitely 
heavy atom} initially in the ground state placed at a node, with the cavity field in a one-quantum coherent 
state. The insets are the ground-level views along the time axis.} % $q=p/\hbar k, \xi=x/\lambda$.}
\label{typical-jump}
\efg

\bfg
%\centerbmp{3in}{2in}{Pe.bmp}
\caption{The evolution of ${\cal P}_e(q)=\sum_n|C_e(n,q)|^2$. Inset shows the temporal behavior of 
$P_e=\int\d q{\cal P}(q)$.}
\label{Pe}
\efg

\bfg
%\centerbmp{3in}{1.75in}{regular-jump-3D.bmp}
\caption{The shapes of the wave packet sitting on a node right after an atomic jump arranged to occur at 
$\tau=1$, $\cdots$, 5, 10 and 20. $q=p/\hbar k, \xi=x/\lambda$. . }
\label{regular-jumps}
\efg

\bfg
%\centerbmp{3in}{2.25in}{far-P_x-1.bmp}
\caption{The ``far-field'' position distributions averaged over 200 atoms at a location approximately $30w$ away 
from the cavity axis. The thinner line is for the case of nonradiative atoms, while the thicker line is for the 
case of radiative atoms with $\gamma/g_0=0.5$.}
\label{fig-exp}
\efg

\bfg
%\centerbmp{3in}{2.25in}{far-P_x-NJ.bmp}
\caption{The shape of the position distribution that would have turned up if all the radiative atoms had gone 
through only the coherent decay (thinner line), compared to the thicker line in FIG.\,\ref{fig-exp}. Each line 
averaged over 200 atoms.}
\label{fig-exp-no-jump}
\efg

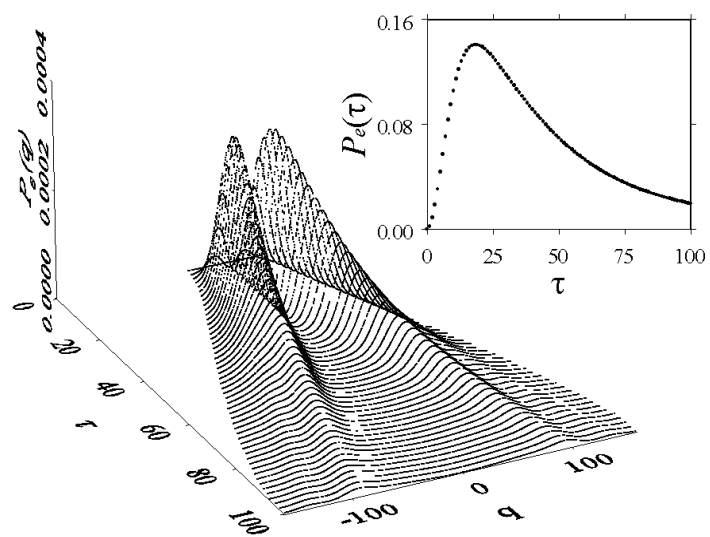
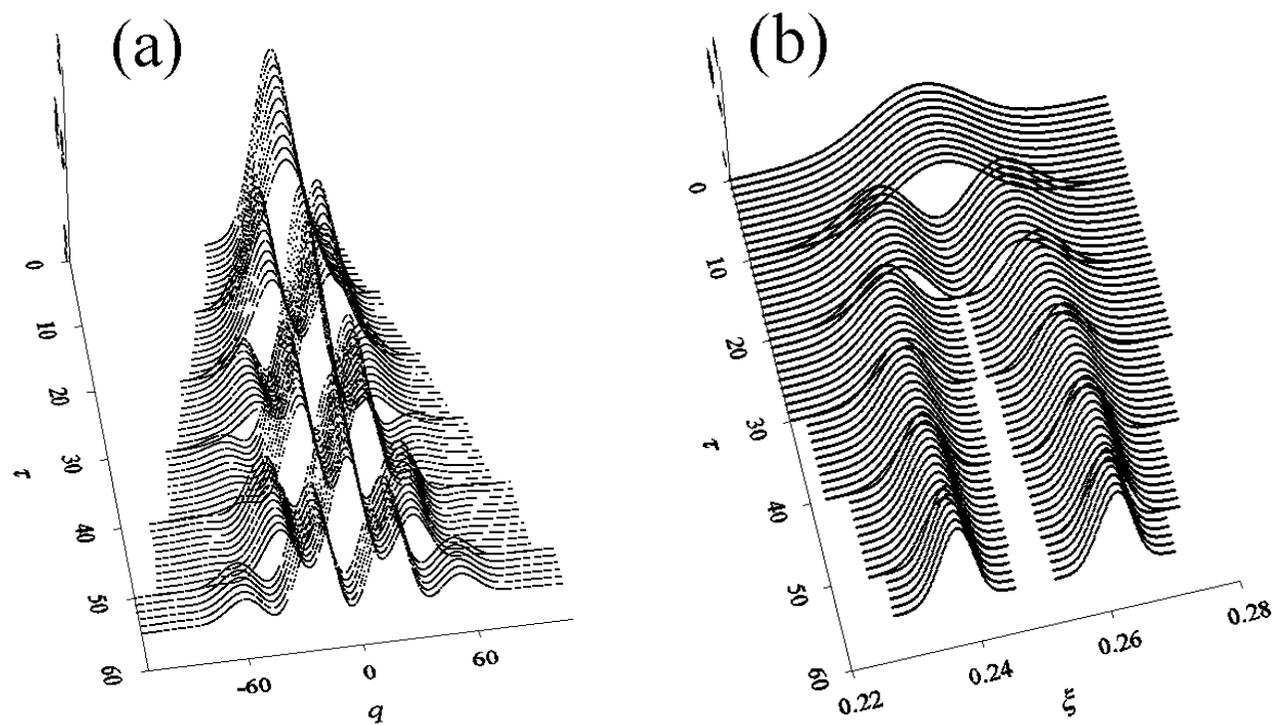
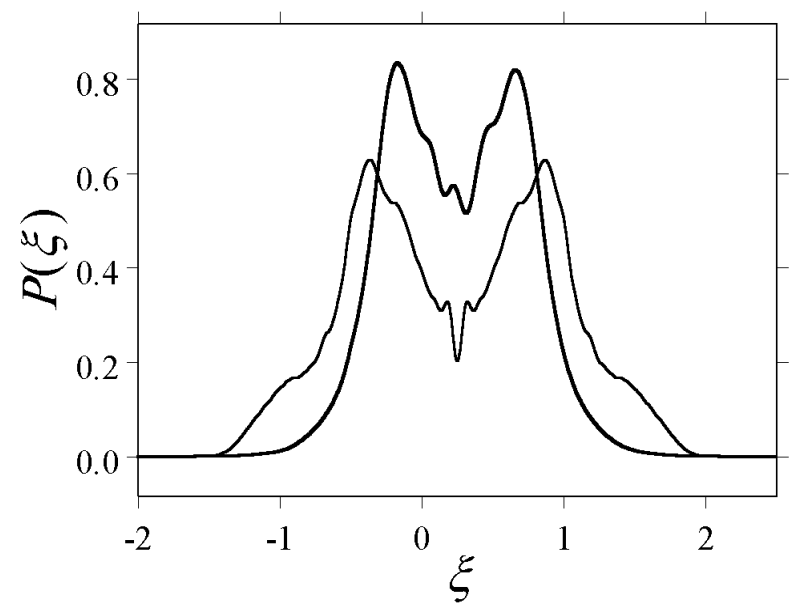
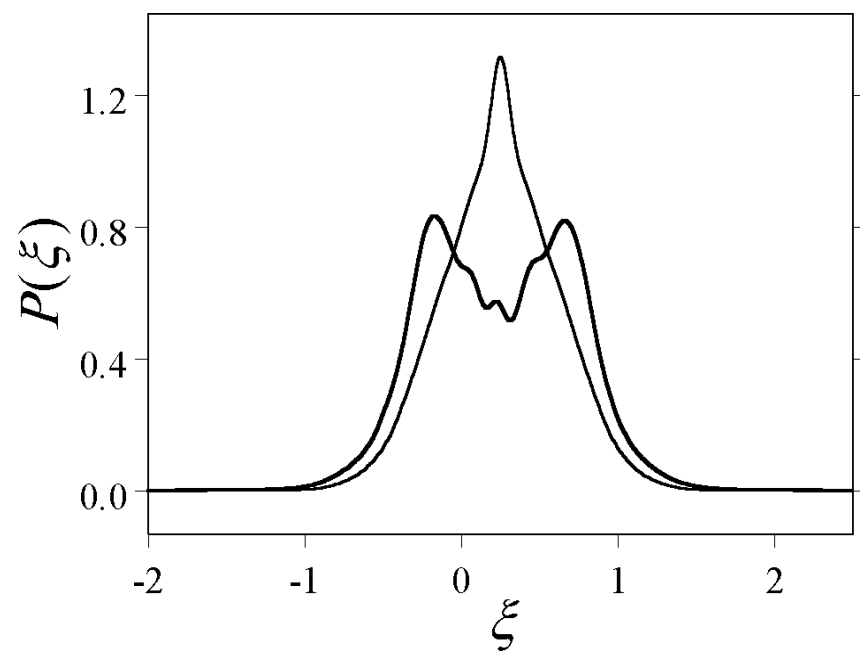
\end{document}